# Evidence of Coulomb interaction induced Lifshitz transition and robust hybrid Weyl semimetal in $T_d$ MoTe$_2$

N. Xu[1,*], Z. W. Wang[2], A. Magrez[3], P. Bugnon[3], H. Berger[3], C. E. Matt[4,5], V. N. Strocov[4], N. C. Plumb[4], M. Radovic[4], E. Pomjakushina[6], K. Conder[6], J. H. Dil[3,4], J. Mesot[3,4,5], R. Yu[2], H. Ding[7,8,*] and M. Shi[4,*]

[1] *Institute of Advanced Studies, Wuhan University, Wuhan 430072, China*

[2] *School of Physics and Technology, Wuhan University, Wuhan 430072, China*

[3] *Institute of Physics, École Polytechnique Fédérale de Lausanne, CH-1015 Lausanne, Switzerland*

[4] *Swiss Light Source, Paul Scherrer Institut, CH-5232 Villigen PSI, Switzerland*

[5] *Laboratory for Solid State Physics, ETH Zürich, CH-8093 Zürich, Switzerland*

[6] *Laboratory for Developments and Methods, Paul Scherrer Institut, CH-5232 Villigen, Switzerland*

[7] *Beijing National Laboratory for Condensed Matter Physics and Institute of Physics, Chinese Academy of Sciences, Beijing 100190, China*

[8] *Collaborative Innovation Center of Quantum Matter, Beijing, China*

\* E-mail: nxu@whu.edu.cn, dingh@iphy.ac.cn, ming.shi@psi.ch

**Using soft x-ray angle-resolved photoemission spectroscopy we probed the bulk electronic structure of $T_d$ MoTe$_2$. We found that on-site Coulomb interaction leads to a Lifshitz transition, which is essential for a precise description of the electronic structure. A hybrid Weyl semimetal state with a pair of energy bands touching at both type-I and type-II Weyl nodes is indicated by comparing the experimental data with theoretical calculations. Unveiling the importance of Coulomb interaction opens up a new route to comprehend the unique properties of MoTe$_2$, and is significant for understanding the interplay between correlation effects, strong spin-orbit coupling and superconductivity in this van der Waals material.**

The $T_d$ phase of MoTe$_2$ belongs to the materials family of two-dimensional transition metal dichalcogenides (2D TMDs), which exhibit a wide variety of novel physical properties and have thus been of increasing interest recently [1-2]. These compounds are formed by layers with strong covalent intra-layer bonds, but adjoining layers are coupled by weak van der Waals' forces. This remarkable anisotropy makes it possible to exfoliate crystals down to a single layer with potential applications for nanoscale electronics [3]. Distinct from its counterpart 2H MoTe$_2$ and other TMDs in hexagonal structure, $T_d$ MoTe$_2$ (in the manuscript only this structural polytype is discussed), forms an orthorhombic lattice with a non-centrosymmetric unit cell containing two MoTe$_2$ layers (Fig. 1a). Furthermore, it shows no semiconducting properties but instead behaves as a semimetal with large and unsaturated magnetoresistance [4]. It has been intensively studied as a potential candidate for a quantum spin Hall insulator in the 2D limit [2]. Recently, density functional theory (DFT) calculations [5-7] predicted bulk MoTe$_2$ to be a promising candidate for a type-II Weyl semimetal (WSM). This prediction is supported by some spectroscopic evidences [8-10]. In contrast to the iso-structural WTe$_2$, MoTe$_2$ undergoes a structural transition from $T_d$ to monoclinic 1T' above 240 K, and an additionally the structural transition from 2H to 1T' can be driven by laser illumination or electrostatic doping [11-14]. Furthermore, among all the $T_d$ phase TMDs, a superconducting transition at ambient pressure has only been observed in MoTe$_2$ [15]. These exotic properties in this 2D material make MoTe$_2$ as a promising material candidate for nano-electronics and topological quantum device applications.

Although many novel features have been revealed in MoTe$_2$, the underlying bulk electronic structure which plays a decisive role in the transport and topological properties, has still not been directly determined from experimental investigations. Due to the large lattice constant along the c axis (Fig. 1a), the Brillouin zone (BZ) is tiny along $k_z$ ($\pi/c$ = 0.226 Å$^{-1}$) (Fig. 1b) and requires an experimental probe having high $k_z$ resolution to resolve the three-dimensional electronic structure. Here, using soft X-ray angle-resolved photoemission spectroscopy (SX-ARPES) with increased $k_z$ resolution [16], we present a comprehensive study of the bulk electronic structure of MoTe$_2$. The determined Fermi surface (FS) and band structure clearly show a periodicity along the $k_z$ direction, indicating coherent hopping of electrons between the layers similar to WTe$_2$ [17]. In contrast to previous studies (reproduced in Fig. 1c) [6-7], we demonstrate that the inclusion of on-site Coulomb interactions in the DFT calculations is essential for the correct description of the band structure and FS topology. This is especially clear for the (N+1)th band (black lines in Fig. 1c-e) around the Y(T) point, which is expected to form type-II Weyl nodes with the Nth band (brown lines in Fig. 1c-d) according to DFT calculations [6-7]. Our SX-ARPES and DFT+U results further suggest that correlation effects can lead to a stable hybrid Weyl semimetal state, where both type-I and type-II Weyl nodes are formed by the same pair of bands near the chemical potential [18].

High quality 1T' single-crystals were produced by chemical vapor transport (CVT). A sealed ampule containing high purity Mo, Te and Iodine, used as transport agent, was placed in a horizontal furnace with the reaction zone at 1000°C and growth zone at 940°C. Large and high quality 1T'-MoTe$_2$ single crystals were obtained after a few weeks. The chemical stoichiometry was measured by x-ray fluorescence with a 30 micrometer spot size. The $T_d$-phase is obtained when cooling the 1T'-MoTe$_2$ below 240 K. Clean (001) surfaces of $T_d$-MoTe$_2$ were prepared for SX-ARPES measurements by cleaving the samples *in situ* in a vacuum better than $5 \times 10^{-11}$ Torr and at a temperature lower than 20 K. Bulk-sensitive SX-ARPES measurements were performed at the Advanced Resonant Spectroscopies beam line (ADRESS) at SLS [19] using a hemispherical electrostatic spectrometer manufactured by the Specs GmbH. Data were collected using circular-polarized light with an overall energy/angular resolution in the range of 50–80 meV/0.1° at T ~ 10 K. Electronic structure calculations were performed based on DFT method. The electron-electron interaction between Mo 4$d$ electrons is simulated through Dudarev's method by setting effective Coulomb interaction $U_{eff} = (U-J)$ in the DFT+U scheme, where the Hund's coupling J = 0.4 eV is taken as the typical value for Mo-related compounds [20-21]. Slightly changing of U-J value is not changing the conclusion of this work. A tight-binding model based on the maximally localized Wannier functions method has been constructed in order to investigate the Weyl points of $T_d$ MoTe$_2$.

Figure 2a shows the ARPES spectra in the $k_x$-$k_z$ plane, acquired from the cleaved (001) surface of MoTe$_2$ with the photon-energy in the range of 350-650 eV. The periodic variation of the Fermi momentum in $k_z$ confirms the bulk origin of the observed electronic states. However, the periodicity in $k_z$ is 4π/c which is twice of that in the bulk BZ along the $k_z$ direction (2π/c). The constant energy map at binding energy ($E_B$) of 0.5 eV (Fig. 2b) shows also a 4π/c periodicity in the $k_z$ direction. This phenomenon is due to matrix element effects in the non-symmorphic crystal symmetry which causes the measured ARPES intensity to depend on the BZ number [22-24]. The 2π/c periodicity can be revealed by measuring at $k_x = 0$ and $k_y = 2π/b$, and the observed valence band dispersion along Γ-Z-Γ can be well reproduced by the DFT calculation with spin-orbit coupling and onsite Coulomb interaction ($U_{eff}$ = 2.4 eV) taken into account (Fig. 2c).

To further explore the bulk electronic structure of MoTe$_2$, we acquired SX-ARPES data in the $k_z = 0$ plane with photon energy hν = 600 eV. The determined bulk FSs consist of a pair of banana-shaped hole-like pockets (α) slightly off the Γ-Y symmetry line and a pair of electron-like pockets (β) located farther from the center in the form of smaller crescent-shape facing the α pockets (Fig. 3a). The DFT (green lines) and DFT+U (red lines) calculations produce similar band dispersions along the Γ-X direction, and both of them are consistent with the experimental results (Fig. 3d). On the other hand, DFT and DFT+U methods give qualitatively different band dispersions along the Γ-Y (Fig. 3e), leading to a change of Fermi surfaces topology near the Y point (Fig. 3c,e). In the DFT calculations (green lines in Fig. 3e), two electron-like bands with tiny spin splitting (γ and δ) cross $E_F$ around the Y(T) point, forming almost 2D FSs with no dispersive feature along the Y-T direction. However, both the γ and δ electron pockets are absent in the ARPES spectra (Fig. 3a and e). The acquired SX-ARPES spectra in a large number of BZ along the Y-T direction show no band within 0.5 eV below $E_F$ (Fig. 3e), thus excluding the existence of the γ and δ pockets. We have also collected SX-ARPES data in a large momentum space, over several Brillouin zones along both in-plane and out-of-plane directions, and found no sign of the electron-like pockets around the Y(T) point. We noticed that the band bottoms of the β and γ bands occur at almost the same energy positions in DFT calculations (Fig. 3d-e). The absence of the γ and δ electron pockets cannot result from any possible doping, because the β pockets are clearly observed in Fig. 3a and d.

To obtain insight into the qualitative difference between the experimentally determined band structure and the one from DFT calculations, we performed electronic structure calculations on MoTe$_2$ using the DFT+U method to include on-site correlations. The most pronounced effect of $U_{eff}$ is that the electron-like bands near the Y(T) point are pushed above $E_F$, resulting in the absence of the γ and δ pockets (Fig. 3b, e), which is consistent with our ARPES results. In normal DFT an unrealistic lattice expansion of 5% is needed to push the states above the Fermi level at the Y(T) point, simultaneously completely altering the dispersion close to the Γ-point (Fig. S1 in Supplementary Material), thus ruling out a lattice effect as a possible explanation for our observations. Further theoretical analysis shows that, for the γ pocket, a Lifshitz transition takes place at a critical value of $U_{eff}$ ~ 1.8 eV (Fig. 1d). We further demonstrate that band structure calculations using the hybrid-functional method (HSE06) or van der Waals density functional method cannot explain the experimentally observed bulk FS configuration changes. The band dispersions and FS topology around the Γ point from different functionals would deviate from the SX-ARPES results if the Liftshiz transition had occurred near the Y point (Figs. S2-S3 in Supplementary Material). The good consistency between the SX-ARPES spectra and the band structure calculated using the DFT+U method with $U_{eff}$ = 2.4 eV strongly indicates that the on-site Coulomb interaction plays a crucial role in influencing both the electronic structure near the chemical potential and the FS topology. The $U_{eff}$ value of 2.4 eV is also the typical value for other Mo-related compounds in previous studies [20-21].

The on-site Coulomb interaction in $MoTe_2$, which was rarely taken into account in previous theoretical consideration, has a striking influence on its topological properties, as summarized in Fig. 4a-d. In the DFT+U calculation with $U_{eff}$ = 2.4 eV and lattice parameters from Ref. [7], the band minimum of the (N+1)th band near the Y(T) point is shifted to above $E_F$. The predicted eight type-II Weyl nodes (open circles in Fig. 4b) in the non-interacting calculations [6], which are located at two different energies in the $k_z$ = 0 plane and are formed by the (N+1)th and Nth bands, transform into four Weyl nodes of a single type (W1) near $E_F$ in the $k_z$ = 0 plane (solid circles in Fig. 4c). Furthermore, the Weyl cone of W1 has a type-I like dispersion, i.e. the Fermi velocities of the crossing bands have opposite sign in any momentum plane that cuts the Weyl cones and contains Weyl points. As shown in Fig. 4e,g,h, the band dispersions passing through W1 from the DFT+U calculation show an overall consistency with the ARPES results along the $k_x$, $k_y$ and $k_z$ directions. We note that W1 is very close to the critical point between type-I and type-II Weyl nodes (Fig.

4e,f), which can, mathematically, be considered as an analogue of a black hole horizon [25]. Interestingly, the Nth and (N+1)th bands from the DFT+U calculation touch again and form eight type-II Weyl nodes (W2) in the $k_z = \pm 0.08$ Å$^{-1}$ plane (open circles in Fig. 4c). The band structure around W2 corresponds to type-II Weyl cone dispersions with a node at 12 meV below $E_F$ (Fig. 4i-j). The good agreement between the ARPES data and the band structure from the DFT+U calculation (Fig. 4i-l) strongly suggests that MoTe$_2$ contains type-II Weyl nodes off the $k_z = 0$ plane, along with type-I Weyl nodes in the $k_z = 0$ plane.

The combined ARPES and DFT+U results suggest a novel topological phase exists in MoTe$_2$, -i.e. a hybrid Weyl semimetal phase induced by electron correlation in which the same conduction and valence bands form both type-I and type-II Weyl nodes near $E_F$ (Fig. 4d). Such a hybrid WSM is predicted to show a unique Landau-level structure and quantum oscillations [18]. We also note that the Weyl nodes with opposite chirality are well separated, which makes the correlated hybrid Weyl semimetal state robust against lattice perturbations. We further theoretically examine the topological properties of MoTe$_2$ using another set of lattice parameters that were experimentally determined at low temperature. A previous DFT calculations suggests that a 0.3% smaller lattice constant *a* could induce a topological phase transition from eight type-II Weyl nodes at two different binding energies (open circles in Fig. 4b) to four type-II Weyl nodes at a single energy (open circles in Fig. 4a) [7]. In contrast, our DFT+U calculation shows that the topological properties of MoTe$_2$ are insensitive to the 0.3% variation of the lattices constant, i.e. the number and types of Weyl nodes are unchanged, and the locations of Weyl nodes are almost at the same binding energies. The hybrid Weyl semimetal state occurs for $U_{eff}$ larger than 2 eV, which is found to reproduce the experimentally observed Lifshitz transition of FS near the Y(T) point, and it persists for values of $U_{eff}$ up to 3 eV. This covers the common $U_{eff}$ values usually used in Mo and other 4d electron systems.

In summary, we demonstrated that electron correlation plays a crucial role in the description of the band structure and FS in $T_d$ MoTe$_2$. Our theoretical analysis shows that a Lifshitz transition occurs upon increasing the on-site $U_{eff}$ over a threshold of ~ 1.8 eV. The bulk band structure measured by SX-ARPES measurements agrees well with the DFT+U calculations using $U_{eff} = 2.4$ eV. Our combined results indicate that a novel topological phase, the hybrid Weyl semimetal state, could emerge from the van der Waals material $MoTe_2$. The hybrid Weyl semimetal state induced by electron correlation is robust against disorder and lattice perturbations. $MoTe_2$ with correlated 4d electrons provides a versatile platform for studying the fundamental physics of novel topological phases and the interplay between spin-obit coupling, Coulomb interaction and superconductivity.

# Figures

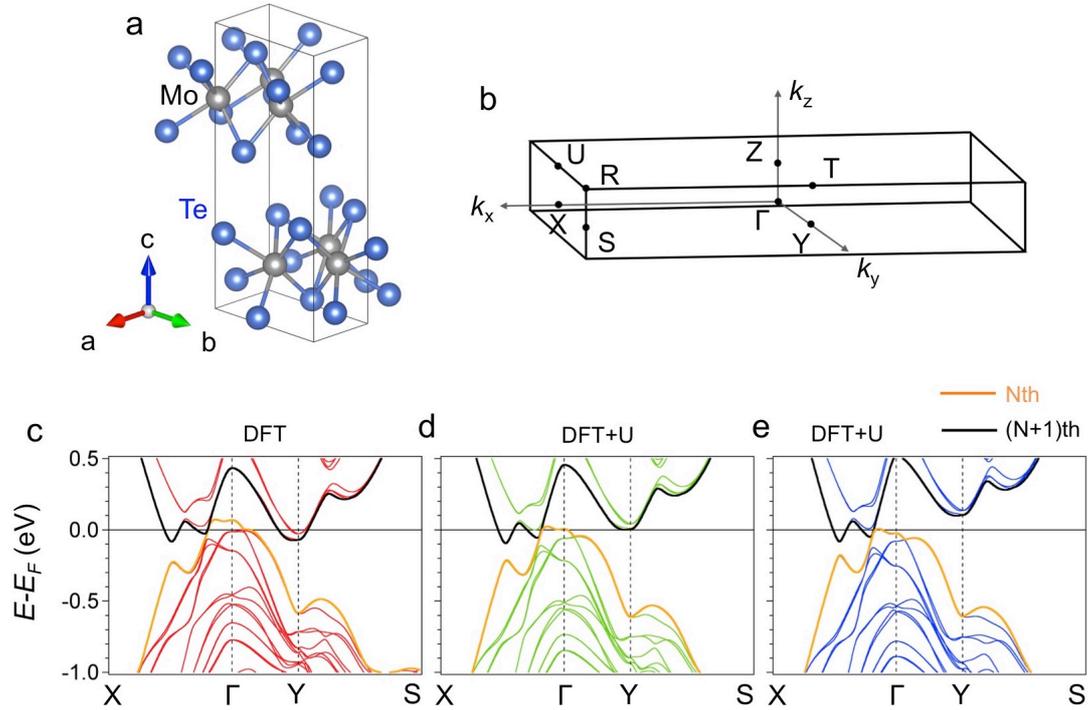

**Figure 1. Correlation induced Lifshitz transition in MoTe$_2$. a,** Crystal structure of the orthorhombic (T$_d$) phase of MoTe$_2$. **b,** Bulk Brillouin zone of MoTe$_2$. **c-e,** Calculated band dispersions using the DFT and DFT+U methods with U$_{eff}$ = 1.8 eV and 3 eV, respectively. The crossings of Nth band (brown line) and (N+1)th band (black line) form type-II Weyl nodes in DFT calculations [6-7].

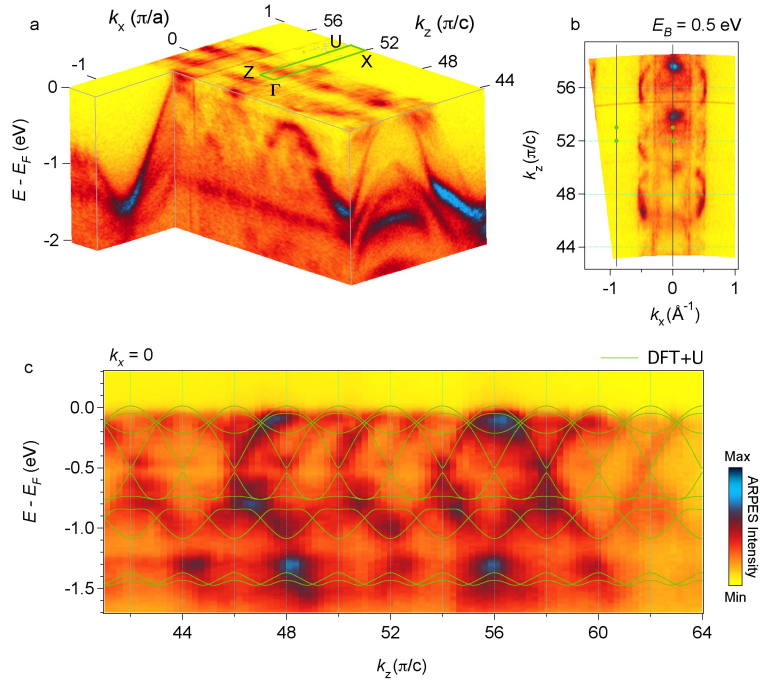

**Figure 2. Bulk electronic structure of MoTe$_2$. a,** Photoemission intensity of the bulk electronic states as a function of energy relative to the Fermi level in the k$_x$-k$_z$ plane, obtained from ARPES measurements in the photon-energy range of 350 – 650 eV. **b,** Constant energy map at 0.5 eV below E$_F$ in the *k$_x$*-*k$_z$* plane. **c,** ARPES intensity plot along the *Γ*-*Z* direction, taken with photon energies covering a *k$_z$* range over 10 BZs. The DFT+U calculated bands are overlaid for a direct comparison.

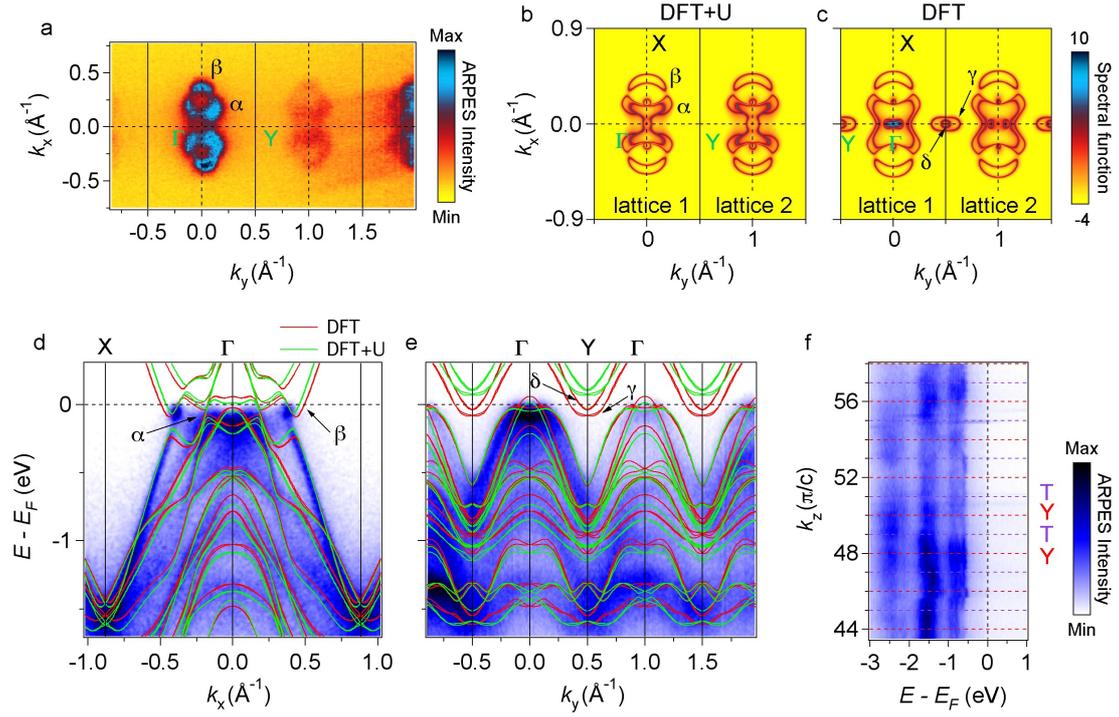

**Figure 3. Effect of Coulomb interaction on electronic structure of MoTe$_2$. a,** FS intensity map in the $k_x$-$k_y$ plane with $k_z = 0$, acquired with hν = 600 eV in the ARPES measurements. **b,** Corresponding spectral functions from bulk band calculations using the DFT+U method. To illustrate the effect of a slight variation of the lattice parameters we plot corresponding FSs for two sets of lattice constants (lattice 1 from Ref. [6] and lattice 2 from Ref. [7]). **c,** The same as **b**, but calculated using the DFT method. **d,** ARPES spectrum near $E_F$ along the $\Gamma$-$X$ direction. The overlaid solid lines are the energy bands calculated by using the DFT+U (green) and DFT (red) methods, respectively. **e,** Same as **d**, but along the $\Gamma$-$Y$ direction. **f,** ARPES spectrum along $T$-$Y$ direction, taken with photon energies covering a $k_z$ range of over 26π/c.

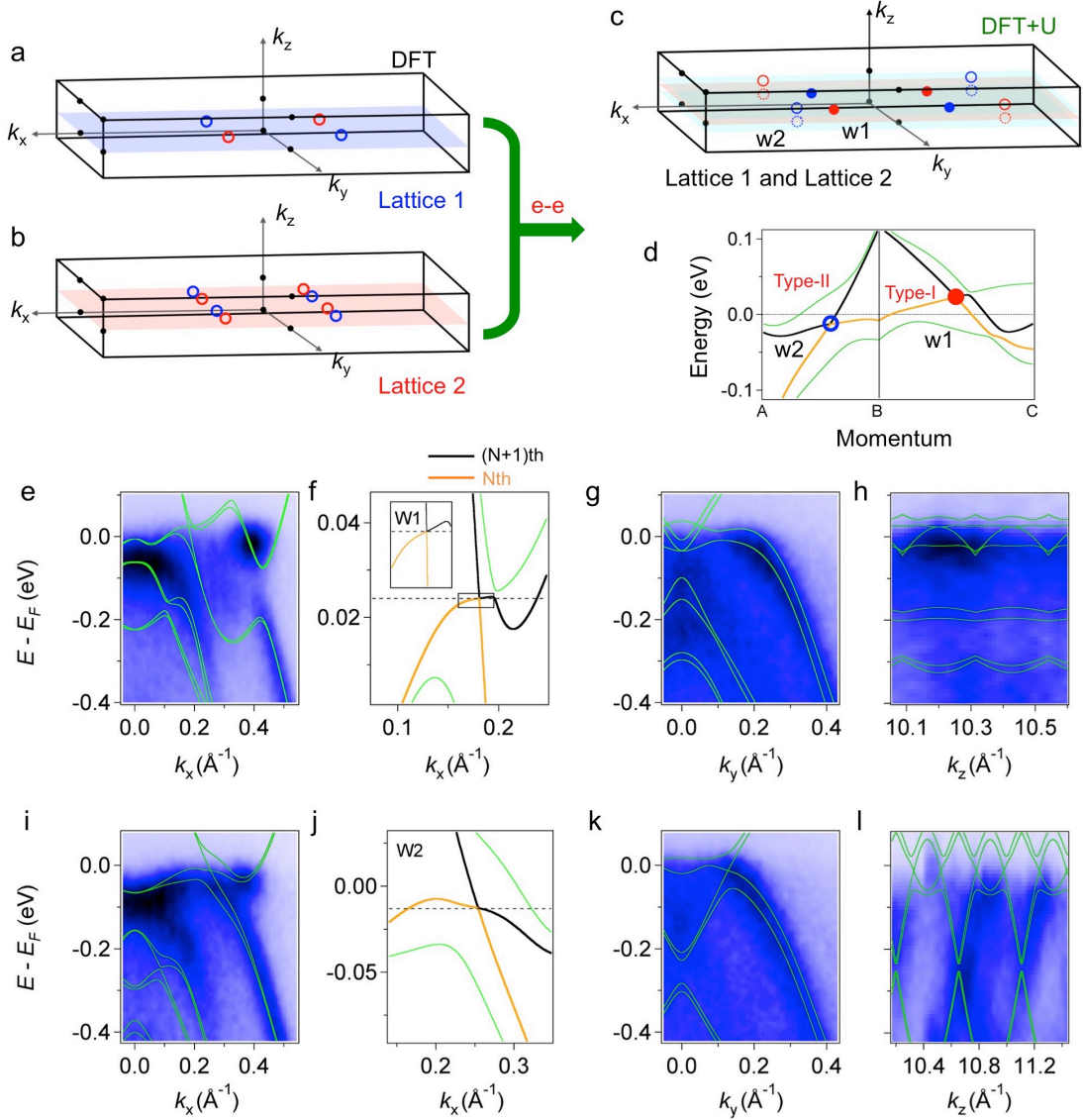

**Figure 4. Evolution of topological properties of MoTe$_2$ induced by Coulomb interaction. a-b,** Illustration of Type-II Weyl nodes in MoTe$_2$ obtained from the DFT calculations using the lattice constants from Ref. [6] and [7], respectively. **c,** Weyl nodes in MoTe$_2$ obtained from DFT+U calculations. Solid and open circles indicate type-I and type-II Weyl nodes, respectively. The blue and red colours represent different chirality. **d,** Band structure passing through both W1 and W2 from DFT+U calculations, which shows the hybrid Weyl semimetal state in MoTe$_2$. The labeled points on the horizontal axis are A = (-0.36, 0.13, 0.08), B = (-0.18, 0.13, 0.08) and C = (-0.18, -0.01, -0.08) in the *k*-space with the unit of Å$^{-1}$. **e,** The ARPES spectrum passing through W1 and along the k$_x$ direction. For comparison, the calculated bands (green curves) are overlaid on top of the experimental data. **f,** Zoomed-in band structure from DFT+U calculations along k$_x$ through W1. **g-h,** Same as **e**, but along

Page 13

the k$_y$ and k$_z$ directions, respectively. **i-l,** Same as **e-h**, but through W2.